\documentclass{article}
\usepackage[a4paper, margin=1in]{geometry}
\usepackage{cite}

\usepackage[pdftex]{graphicx}
\usepackage[cmex10]{amsmath}
\usepackage[caption=false,font=footnotesize]{subfig}

\usepackage{newclude}
\usepackage{changes}
\usepackage{booktabs}
\usepackage{siunitx}

\usepackage{tikz}
\usepackage[european]{circuitikz}
\usetikzlibrary{positioning}
\usetikzlibrary{fit}
\usetikzlibrary{backgrounds}
\usetikzlibrary{calc}
\usetikzlibrary{shapes}
\usetikzlibrary{shadows}
\usetikzlibrary{chains}
\usetikzlibrary{trees}
\usetikzlibrary{arrows}
\usetikzlibrary{intersections}
\usetikzlibrary{mindmap}
\usetikzlibrary{decorations.text}
\usetikzlibrary{spy}
\usetikzlibrary{patterns}
\usetikzlibrary{shapes.geometric, arrows}

\newcommand{\footremember}[2]{%
    \footnote{#2}
    \newcounter{#1}
    \setcounter{#1}{\value{footnote}}%
}
\newcommand{\footrecall}[1]{%
    \footnotemark[\value{#1}]%
}

\begin{document}
\title{OPF-Based Optimal Power System Network Restoration Considering Frequency Dynamics}

\author{
Dawn Virginillo,\footremember{epfl}{D. Virginillo and M. Paolone are with the Distributed Electrical Systems Laboratory, \'{E}cole Polytechnique F\'{e}d\'{e}rale de Lausanne, Lausanne, Switzerland (e-mail: \{dawn.virginillo; mario.paolone\}@epfl.ch).}\footremember{corr}{Corresponding author.} Asja Dervi\v{s}kadi\'{c},\footremember{sg}{A. Dervi\v{s}kadi\'{c} is with Swissgrid AG, Aarau, Switzerland (e-mail: asja.derviskadic@swissgrid.ch).} and Mario Paolone\footrecall{epfl}
}

\maketitle

\begin{abstract}
Due to recent blackout and system split incidents in power grids worldwide, as well as increased system complexity in view of the energy transition, there has been increasing interest in re-evaluating existing Power System Restoration (PSR) plans.  In restoration scenarios, due to low island inertia, it is necessary to ensure not only the static, but also the dynamic stability of the system.  In this paper, we pose and solve a formulation of the optimal PSR problem including frequency dynamics.  We validate the switching constraints for global optimality within a static version of the formulation using a brute-force tree search method. We apply the dynamic problem formulation to the IEEE 9-Bus model, and show that the optimal switching sequence using the static formulation would violate dynamic constraints, illustrating the importance of dynamic considerations in PSR planning.

\end{abstract}

\section{Introduction}

Owing to a number of grid incidents in the last few years, and in particular, the wide-reaching impacts of the April 2025 blackout of the transmission networks (TNs) of Spain and Portugal~\cite{spain-bo-report}, interest in Power System Restoration (PSR) studies has been growing.  The goal of PSR planning is to design procedures which minimize the time from initial blackout to full re-energization of the system, without additional interruptions. Furthermore, there has been increasing interest in the re-evaluation of existing restoration plans due to the proliferation of Distributed Energy Resources (DERs).  While some work has focused on the goal of incorporating DERs into existing restoration processes~\cite{Zhao2021}, recent incidents have also highlighted the potential risks associated with DERs in the PSR context in the absence of dedicated mitigation plans~\cite{Sanniti2023}.

Generally, real restoration processes are divided into three phases: \textit{black start}, where Black Start Units (BSUs) are energized without external supporting voltage, \textit{backbone construction}, where Non-Black Start Units (NBSUs) and known or controllable loads are picked up to build the electrical island, and \textit{load restoration}, where the remaining distribution grid loads are energized~\cite{Paolone-book}.  In the backbone buildup phase, it is especially critical to study the network dynamics in detail, as low levels of connected inertia imply that the system is sensitive to disturbances~\cite{Adibi-TF}.  Furthermore, a sub-optimal backbone buildup strategy can imply a larger time requirement for the entire restoration process.  Existing PSR processes defined by TSOs often use heuristics to define load pickup strategies (e.g., defining the desired load pickups as a certain fraction of the connected generation)~\cite{Adibi-TF} and are commonly based on thorough simulations of potentially sub-optimal backbone construction paths; implementation of advanced tools based on optimal PSR remains uncommon~\cite{cigre-restore-2017}.

The goal of this work is to investigate optimal PSR strategies for backbone construction using Mixed-Integer Programming (MIP) techniques, inspired by literature on Transient Stability-Constrained Optimal Power Flow (TSC-OPF)~\cite{Alvarez2024,Zhao_Bialek2016}.  The developed formulation defines a fixed amount of time allotted for this phase of PSR, and maximizes the load served in terms of energy over the restoration horizon.  Although some works propose cost-based objectives (e.g.,~\cite{Chou2013,Jiang2023}), we argue that the core objective of a TSO should be the energy served, as normal power markets should not be operational in this special operational situation.

In this paper, we propose a formulation for the optimal PSR problem considering frequency dynamics, which we call dynamic Optimal Power Flow for Restoration (dynOPF-R).  To keep the problem tractable, power flow constraints use a DC OPF formulation.  Dynamic constraints are implemented using numerical integration, incorporating droop control of generators and the swing equation for computation of frequency dynamics~\cite{Kundur}.  As the problem is combinatorial, the developed MIP formulation has both linear and bilinear constraints, where the latter involves multiplication of binary variables (i.e., switching variables) by continuous ones (i.e., frequency and power output of generators).  As recent developments in MIP solvers have enabled solution of problems with such bilinear constraints to global optimality~\cite{gurobi-nonlin}, the developed formulation can be solved by existing commercial solvers using, for example, branch and bound methods.  In comparison with existing works, the proposed formulation includes the entire transient, enabling validation against existing dynamic models, and is solved in a single stage.

The formulation without dynamics is also presented in this paper, which enables an examination of the impact of including dynamic constraints on the optimal restoration path.  Validation of the switching constraints is conducted by brute force on the static formulation, using an implementation of in-order transversal (depth first) tree search, to verify global optimality of the restoration path.  The dynamic equations are validated against a detailed dynamic model.  The developed dynamic formulation is implemented on the standard IEEE 9-Bus model, ensuring replicability with a focus on transmission-level networks.  It is further demonstrated that the optimal switching path computed by the static formulation would result in frequency violations, illustrating the importance of considering dynamic constraints in restoration planning processes.

\noindent The contributions of the paper are as follows:
\begin{itemize}
   \item Presentation and validation of an efficient algorithm for optimization of the energization order in PSR operations based on DC Optimal Power Flow (OPF), and
   \item Proposal of an optimal PSR problem formulation considering frequency dynamics.
\end{itemize}

The paper is organized as follows.  In Section~\ref{chap:sota}, we provide a literature review on the optimal PSR problem.  In Section~\ref{chap:static}, we present and validate the static formulation of the problem.  We present the proposed dynamic formulation of the model in Section~\ref{chap:dynopf-r}, and show results in Section~\ref{chap:results}.  Finally, we draw conclusions in Section~\ref{chap:conc}.

\section{State of the Art}\label{chap:sota}

Formulations for the optimal PSR problem have been widely reported in the literature, with each work including a different range of phenomena in its objectives and constraints. Various formulations have been proposed which consider only static power flow constraints, such as in~\cite{Patsakis2018}, which uses active and reactive power balances. Formulations of the power flow constraints using Second Order Cone Programming (SOCP)~\cite{ZhaoShen2021} and linearized AC power flow~\cite{Liu2020} have also been proposed.  In~\cite{Viana2013}, the formulation which computes the maximum load which can be picked up, considering the instant directly after energization and the steady-state system characteristics; however, frequency dynamics are not considered.  

Some recent works have considered the impact of system dynamics in the optimal PSR problem, which are outlined as follows.  Multiple works have proposed the use of offline frequency dynamics models to compute the maximum load that can be picked up, followed by a static optimal PSR formulation~\cite{Gholami2017, Golshani2019, Sun2022}.  In~\cite{LIAO2019265}, a two-stage iterative optimization problem formulation is presented, where the first stage optimizes the switching sequence and the second computes system dynamics.  In~\cite{Zhao2021}, the optimal restoration path is computed considering TSO-DSO coordination, using a simplified model for frequency dynamics which considers the frequency as directly proportional to the power imbalance.  In~\cite{Jiang2023}, the start-up sequence of generators is studied considering the penetration of RES, which incorporates frequency dynamics using a linearized form of the swing equation.  However, to enable analytical computation of the frequency nadir, generators' production in response to a load step is simplified and modelled as a piecewise linear ramp. Finally, a similar method of considering frequency dynamics is used in~\cite{Qin2025}, where a bilevel formulation is proposed which optimizes the restoration path in the presence of stochastic RES. In~\cite{Hijazi2015}, a detailed frequency model based on numerical integration is considered, but it is used as a dynamic stability check rather than integrated in the main optimization problem.

Solution approaches for presented formulations can generally be divided into two categories: heuristic methods and MIP, where the latter is usually solved using commercially available tools with branch-and-bound algorithms.  

Some examples of heuristic approaches to the optimal PSR problem are outlined as follows.  In~\cite{Chou2013}, a swarm intelligence approach is applied to optimize the restoration plan of Taiwan.  In~\cite{Sun_Terzija_2019}, a genetic algorithm is used to solve a formulation of the optimal PSR problem with multiple objectives. In~\cite{Hsiao2004}, a method for topological reconfiguration using fuzzy sets is proposed.  While this category of solution methods often offers better performance with increasing problem scale and can handle non-convex formulations, convergence to the global optimum cannot be guaranteed.

It is known that the problem is highly combinatorial, and can therefore be difficult to solve using MIP formulations; some works have addressed this issue directly.  In~\cite{Patsakis2018}, scalability issues are shown when scaling up the problem to the IEEE 118 Bus test system, and a heuristic is proposed to address this issue. In~\cite{Xie2024}, a bulk model is used for frequency dynamics to ease the computational burden. Multiple works use multi-stage formulations for computational efficiency; in such cases, it is common to have the main problem contain the switching constraints, while the sub-problem optimizes the operating points (e.g.,~\cite{Gholami2017,Qin2025}).  However, it is noted that most optimal PSR problems are intended for offline use; therefore, computational tractability is a more pertinent goal than efficiency of solution.

Further, it is noted that many works focus on modelling the impact of stochastic RES in the early stages of the PSR process (e.g.~\cite{Liu2020},~\cite{Qin2025},~\cite{Qiu2023}), using methods such as wait-and-see bilevel optimization approaches and chance constraints.  However, we assert that such approaches are impractical for real-world restoration operations, due to the inherent need of the TSO to maximize the success of the restoration process and thus minimize risks stemming from stochastic generation.  This point is illustrated in the restoration strategy of one European TSO~\cite{ELIAresto}, which plans to restore the grid at 51 Hz in the early stages to avoid stochastic injections of power from photovoltaic power plants, and also by the 2025 post-blackout restoration of the Iberian power system, during which the relevant authorities only allowed production from renewable generation once the restoration process was sufficiently advanced~\cite{spain-bo-report}. 

Owing to recent advances in computational power and optimization methods, MIP approaches have gained increasing interest to solve the TSC-OPF problem.  In~\cite{Alvarez2024}, a nonlinear AC TSC-OPF is implemented using numerical integration to assess security issues on microgrids with stochastic RES. A similar approach is applied in~\cite{Zhao_Bialek2016} with a DC approximation, in which a framework for primary frequency control is proposed. In works on scheduling problems with dynamic constraints, the use of frequency and/or RoCoF envelopes~\cite{Alic2024} and Bernstein polynomials to approximate system dynamics~\cite{Rajabdorri2024} have been proposed. In this paper, we propose a mixed-integer formulation for the optimal PSR problem, inspired by literature on TSC-OPF.  In contrast with existing works on the optimal PSR problem, our approach computes the optimal switching sequence and frequency dynamic in an integrated formulation (single-stage), and handles generator and frequency dynamics in a physics-informed manner, enabling validation against existing dynamic models.  Additionally, we handle generator pickups in a manner which is integrable between the switching and dynamic constraints, using a similar method as has been applied for island synchronization~\cite{Maharjan2025}, enabling wholistic optimization of the dynamic restoration process.

\section{Optimal Power Flow for Restoration: OPF-R}\label{chap:static}

In this section, a static formulation of the optimal restoration problem (OPF-R) is presented, which is similar to some formulations which have been proposed in the literature (i.e.,~\cite{Qin2025,Golshani2019}).  Thus, the purpose of this section is to present the structure of the PSR switching constraints and validate the global optimality of the implemented formulation.  The optimality validation is conducted by brute force tree search.  Analysis of the static formulation results facilitates comparison with the dynamic formulation (DynOPF-R) in the next section. 

\subsection{Formulation of the Problem}
The static problem formulation is based on DC OPF, which neglects voltage and reactive power constraints by construction.  With this assumption, the flow constraints are kept linear, avoiding quadratic voltage constraints. The objective of this approximation is to provide a reasonable grid model which is computationally tractable when integrated with binary variables and frequency dynamics.  Future work will focus on incorporating more accurate grid constraints. 

\noindent First, the formulation parameters are defined as follows:
\begin{itemize}
    \item $\mathbf{I}_d, \mathbf{I}_g, \mathbf{I}_l, \mathbf{I}_w$ are the incidence matrices of loads, generators, lines, and transformers, respectively, and $\mathbf{i}_i$ is the incidence of an individual component.  Incidence matrices are of dimensions $n_b \times n_i$, where $n_b$ is the number of buses and $n_i$ is the number of components.
    \item $P_d$ is the nominal power of load $d$, expressed in matrix form as $\mathbf{P}_d$.
    \item $Y_i$ is the scalar admittance of component $i$.
    \item $\underline{P_g}$ and $\overline{P_g}$ are the upper and lower limits of active power generation for generator $g$.
    \item $\underline{P_l}$ and $\overline{P_l}$ are the upper and lower limits of active power flow on line $l$, and analogously for transformers.
    \item $\mathcal{B}$ is the set of BSUs and $\mathcal{N}$ is the set of NBSUs.
    \item $\mathcal{D}$, $\mathcal{L}$, $\mathcal{W}$ are the sets of loads, lines, and transformers in the system, respectively.
    \item $\mathcal{T}$ is the set of timesteps, such that $\mathcal{T} = [1,2,...,T_{max}]$, and $\mathcal{T}'$ is the set of timesteps excluding the initialization, such that $\mathcal{T}' = [2,3,...,T_{max}]$.
\end{itemize}

Further, the problem variables are listed as follows:
\begin{itemize}
    \item $\mathbf{S}_d, \mathbf{S}_g, \mathbf{S}_l, \mathbf{S}_w$ are the binary switching variables of loads, generators, lines, and transformers, respectively, expressed as matrices with dimension $n_i \times n_t$, where $n_t$ is the number of timesteps in the model.  Switching variables are expressed in scalar form as $S_{i,t}$.
    \item $P_g$ is the active power produced by generator $g$, expressed in matrix form as $\mathbf{P}_g$.
    \item $\mathbf{P}_l,\mathbf{P}_w$ are the matrices of active power flows on lines and transformers, respectively.
    \item $\boldsymbol{\theta}$ is the matrix of bus voltage angles.
    \item $\mathbf{S}_b$ the energization of buses (integer variable), defined by the number of live connections to each bus.
\end{itemize}

The problem objective is defined as the maximal energy served over the time horizon of the restoration process.  For ease of implementation, the time allotted for the restoration process is fixed, and it is assumed that the loads are static (i.e., controlled only by binary variables) and follow constant power characteristics.  The mathematical formulation of the problem (in per unit) is as follows:
\begin{subequations}
    \begin{alignat}{4}
        &\max_{S,P_g,P_l,P_w,\theta} \quad  & & \sum_{d,t} S_{d,t} P_d    \label{eq:obj} \\
        &\text{s.t.} & &  \begin{aligned} \mathbf{I}_d (\mathbf{S}_d \circ \mathbf{P}_d) - \mathbf{I}_g (\mathbf{S}_g \circ \mathbf{P}_g) \\ + \mathbf{I}_l \mathbf{P}_{l} + \mathbf{I}_w \mathbf{P}_{w} = 0 \end{aligned} \label{c:powerbal}\\
        &                  & &  \mathbf{p}_{i} = Y_i \mathbf{s}_{i} \circ \mathbf{i}_i \boldsymbol{\theta} \quad \forall i \in [\mathcal{L}, \mathcal{W}]  \label{c:branchp} \\   
        &                  & &  S_{i,0} = 0  \quad \forall i \in [\mathcal{D}, \mathcal{L}, \mathcal{W}] \label{c:0init}\\
        &                  & &  S_{g,0} = 0 \quad \forall g \in \mathcal{N}  \label{c:nbsuinit}\\
        &                  & &  S_{g,0} = 1 \quad \forall g \in \mathcal{B} \label{c:bsuinit}\\
        &                  & &  \begin{aligned} S_{i,t}-S_{i,t-1} \geq 0 \quad \forall i \in [\mathcal{B}, \mathcal{N}, \\ \mathcal{D}, \mathcal{L}, \mathcal{W}], t \in \mathcal{T}'  \end{aligned} \label{c:cont-energ}\\
        &                  & &  \begin{aligned} \sum_{i \in [\mathcal{N}, \mathcal{D}, \mathcal{L},\mathcal{W}]} S_{i,{t}}-S_{i,t-1} \leq 1  \quad \forall t \in \mathcal{T}' \end{aligned} \label{c:1atatime} \\
        &                  & &  \mathbf{S}_{b} = \sum_{i \in \mathcal{B}, \mathcal{N}, \mathcal{D}} \mathbf{I}_i \mathbf{S}_i + \sum_{j \in \mathcal{L}, \mathcal{W}} |\mathbf{I}_j| \mathbf{S}_j   \label{c:busdef} \\
        &                  & &  S_{d,t}  \leq \mathbf{i}_d \mathbf{s}_{b,{t-1}} \quad \forall d \in \mathcal{D}, t \in \mathcal{T}' \label{c:bus_loads}\\   
        &                  & &  S_{g,t}  \leq \mathbf{i}_g \mathbf{s}_{b,{t-1}} \quad \forall g \in \mathcal{N}, t \in \mathcal{T}'  \label{c:bus_gens} \\ 
        &                  & &  \begin{aligned} S_{i,t} \leq \mathbf{i}_i^+ \mathbf{s}_{b,{t-1}} + \mathbf{i}_i^- \mathbf{s}_{b,{t-1}}  \\ \forall i \in \mathcal{L}, \mathcal{W}, t \in \mathcal{T}' \end{aligned} \label{c:bus_lines} \\
        &                  & &  \underline{P_g} \leq P_{g,t} \leq \overline{P_g}  \quad \forall g \in [\mathcal{B},\mathcal{N}], t \in \mathcal{T}\label{c:genlims} \\
        &                  & &  \underline{P_l} \leq P_{l,t} \leq \overline{P_l}  \quad \forall l \in \mathcal{L}, t \in \mathcal{T} \label{c:linelims} \\
        &                  & &  \underline{P_w} \leq P_{w,t} \leq \overline{P_w}  \quad \forall w \in \mathcal{W}, t \in \mathcal{T}\label{c:trafolims}\\
        &                  & &  -\pi \leq \boldsymbol{\theta} \leq \pi  \label{c:thetalims}\\
        &                  & &  \theta_{slack,t} = 0 \quad \forall t \in \mathcal{T} \label{c:theta_slack}
    \end{alignat}
\end{subequations}

Of the problem variables, it is noted that branch flows $\mathbf{P}_l$, $\mathbf{P}_w$ and bus energization $\mathbf{S}_b$ are dependent variables and deterministically constrained.  The constraints can be divided into four parts: power balance (OPF) constraints, initialization, switching constraints, and component limits.

Constraints~(\ref{c:powerbal}) and~(\ref{c:branchp}) make up the OPF constraints.  Constraint~(\ref{c:powerbal}) imposes a zero power balance per bus at each timestep\footnote{The symbol $\circ$ represents the matrix Hadamard product.}, while~(\ref{c:branchp}) computes the power flows on lines and transformers (expressed in vector form).  These constraints are mixed-integer bilinear, which are by definition non-convex.  However, as commonly available optimization solvers have the capability to perform convex relaxations on such constraints, using techniques such as McCormick envelopes~\cite{gurobi-nonlin}, they are implemented in the model as written.

The initialization is specified as follows. Constraint~(\ref{c:0init}) specifies that all lines, loads, and transformers are initially switched off.  Constraint~(\ref{c:nbsuinit}) specifies that NBSUs are intially de-energized, while Constraint~(\ref{c:bsuinit}) specifies that BSUs are initially energized.  Such an initialization represents a complete black start scenario, which can be adjusted by the modeler to solve the problem for partial blackout scenarios.

The switching constraints are defined as follows. First, Constraint~(\ref{c:cont-energ}) requires that a component must stay energized once it is switched on.  Constraint~(\ref{c:1atatime}) specifies that one component can be switched on per timestep.  The integer matrix variable $\mathbf{S}_b$ is used to denote the energization of a busbar, defined as the number of live connections to the bus, as specified in Constraint~(\ref{c:busdef}).  Constraints~(\ref{c:bus_loads})--(\ref{c:bus_lines}) specify that elements can only be switched on if an adjacent bus is live (adapted from~\cite{Qin2025}).

Finally, Constraints~(\ref{c:genlims}),~(\ref{c:linelims}), and~(\ref{c:trafolims}) impose active power limits on generators, lines, and transformers, respectively.  Constraint~(\ref{c:thetalims}) constrains the bus angles to reasonable values, while Constraint~(\ref{c:theta_slack}) imposes a zero slack bus angle.  The latter two constraints are convenient to avoid multiplicity of identical solutions.

\subsection{Validation on the 9-Bus Model}

\begin{figure}[t]
    \centering
    \resizebox{8.8cm}{4.9cm}{
    \begin{tikzpicture}[x=3cm,y=3cm,font=\large]
        \coordinate                    (BUS2) at (0.25,0);
        \coordinate                    (BUS7) at (1,0);
        \coordinate                    (BUS8) at (2,0);
        \coordinate                    (BUS9) at (3,0);
        \coordinate                    (BUS3) at (3.75,0);
        \coordinate                    (BUS5) at (1.5,-0.4);
        \coordinate                    (BUS6) at (2.5,-0.4);
        \coordinate                    (BUS4) at (2,-0.9);
        \coordinate                    (BUS1) at (2,-1.6);
        \coordinate                    (BUS7_2) at (1,-0.1);
        \coordinate                    (BUS8_2) at (2,-0.1);
        \coordinate                    (BUS9_2) at (3,-0.1);
        \coordinate                    (BUS4_2) at (1.9,-0.9);
        \coordinate                    (BUS4_3) at (2.1,-0.9);
        \coordinate                    (BUS5_2) at (1.4,-0.4);
        \coordinate                    (BUS5_3) at (1.6,-0.4);
        \coordinate                    (BUS6_2) at (2.6,-0.4);
        \coordinate                    (BUS6_3) at (2.4,-0.4);
        \coordinate                    (GEN1) at (2,-2.1);
        \coordinate                    (GEN2) at (-0.25,0);
        \coordinate                    (GEN3) at (4.25,0);
    
        % lines
        \draw (BUS7) to[short,*-*] (BUS8) to[short,*-*] (BUS9);
        \draw (BUS7_2) to[short,*-] (1.4,-0.1) to[short,-*] (BUS5_2);
        \draw (BUS9_2) to[short,*-] (2.6,-0.1) to[short,-*] (BUS6_2);
        \draw (BUS5_3) to[short,*-] (1.6,-0.5) -- (1.9,-0.8) to[short,-*] (BUS4_2);
        \draw (BUS6_3) to[short,*-] (2.4,-0.5) -- (2.1,-0.8) to[short,-*] (BUS4_3);

        % Line switches
        \draw (BUS1) ++ (0,0.1) node[squarepole]{};
        \draw (BUS1) ++ (0,-0.1) node[squarepole]{};
        \draw (BUS7) ++ (0.1,0) node[squarepole]{};
        \draw (BUS7) ++ (-0.1,0) node[squarepole]{};
        \draw (BUS8) ++ (0.1,0) node[squarepole]{};
        \draw (BUS8) ++ (-0.1,0) node[squarepole]{};
        \draw (BUS9) ++ (0.1,0) node[squarepole]{};
        \draw (BUS9) ++ (-0.1,0) node[squarepole]{};
        \draw (BUS2) ++ (0.1,0) node[squarepole]{};
        \draw (BUS2) ++ (-0.1,0) node[squarepole]{};
        \draw (BUS3) ++ (0.1,0) node[squarepole]{};
        \draw (BUS3) ++ (-0.1,0) node[squarepole]{};
        \draw (BUS7_2) ++ (0.1,0) node[squarepole]{};
        \draw (BUS9_2) ++ (-0.1,0) node[squarepole]{};
        \draw (BUS5_2) ++ (0,0.1) node[squarepole]{};
        \draw (BUS5_3) ++ (0,-0.1) node[squarepole]{};
        \draw (BUS6_2) ++ (0,0.1) node[squarepole]{};
        \draw (BUS6_3) ++ (0,-0.1) node[squarepole]{};
        \draw (BUS4) ++ (0,-0.1) node[squarepole]{};
        \draw (BUS4_2) ++ (0,0.1) node[squarepole]{};
        \draw (BUS4_3) ++ (0,0.1) node[squarepole]{};
        
        % busbars
        \draw [thick] (0.25,-0.15) -- (0.25,0.15) node[above]{Bus 2};
        \draw [thick] (1,-0.15) -- (1,0.15) node[above]{Bus 7};
        \draw [thick] (2,-0.15) -- (2,0.15) node[above]{Bus 8};
        \draw [thick] (3,-0.15) -- (3,0.15) node[above]{Bus 9};
        \draw [thick] (3.75,-0.15) -- (3.75,0.15) node[above]{Bus 3};
        \draw [thick] (1.35,-0.4) node[left]{Bus 5} -- (1.65,-0.4);
        \draw [thick] (2.35,-0.4) -- (2.65,-0.4) node[right]{Bus 6};
        \draw [thick] (1.85,-0.9) -- (2.15,-0.9) node[right]{Bus 4};
        \draw [thick] (1.85,-1.6) -- (2.15,-1.6) node[right]{Bus 1};
        
        % voltage sources
        \draw (GEN2) node[vsourcesinshape, rotate=90, anchor=north](S){} (S.south) to[short,-*] (BUS2);
        \draw (S.west) node[below]{$SM_2$};
        \draw (GEN3) node[vsourcesinshape, rotate=90, anchor=south](S){} (S.north) to[short,-*] (BUS3);
        \draw (S.west) node[below]{$SM_3$};
        \draw (GEN1) node[vsourcesinshape, rotate=90, anchor=west](S){} (S.east) to[short,-*] (BUS1);
        \draw (GEN1) node[below]{$SM_1$};

        % transformers
        \draw (BUS2) to [oosourcetrans] (BUS7);
        \draw (BUS9) to [oosourcetrans] (BUS3);
        \draw (BUS4) to [oosourcetrans,*-] (BUS1);
        
        % loads
        \draw [-stealth](BUS5) to[short,*-] (1.5,-0.7) node[left]{$L_5$};
        \draw [-stealth](BUS6) to[short,*-] (2.5,-0.7) node[right]{$L_6$};
        \draw [-stealth](BUS8_2) to[short,*-] (2.1,-0.1) -- (2.1,-0.3) node[right]{$L_8$};
        \draw (BUS8_2) ++ (0.1,0) node[squarepole]{};
        \draw (BUS6) ++ (0,-0.1) node[squarepole]{};
        \draw (BUS5) ++ (0,-0.1) node[squarepole]{};
    \end{tikzpicture}
    }
    \caption{IEEE 9-bus model, adapted from~\cite{pscontrolstab}.}
    \label{fig:9busmodel}
\end{figure}
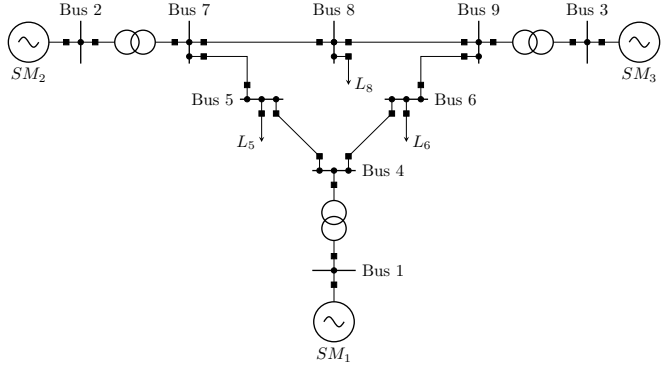

The static PSR formulation presented in the previous section is first validated on the IEEE 9-Bus model~\cite{psanalysis}, depicted in Fig.~\ref{fig:9busmodel}.  The base power $A_\textrm{base}$ is chosen as 200 MW, equivalent to the active power limits of the generators.  Further, the loads $L_5$, $L_6$, and $L_8$, of magnitudes 125 MW, 90 MW, and 100 MW, are split into three equally sized load steps of 41.67 MW, 30 MW, and 33.33 MW.  This modification creates load steps which are more representative of a real restoration process on a high-voltage transmission network.

The computational environment for simulations is described as follows.  Optimization problems are formulated using Yalmip~\cite{yalmip} for Matlab~\cite{MATLAB:2024} and solved using Gurobi v12~\cite{gurobi} via a branch-and-bound algorithm. All simulations \textcolor{blue}{in this section} are conducted on a desktop computer with a 3.2 GHz Intel processor and 32 GB of RAM.

The solution of the OPF-R formulation for the described 9-Bus model with 21 timesteps (20 decision steps plus the defined initial condition) is given in Table~\ref{tab:9bus-static}.  For ease of presentation, each component to be energized is represented by its single-letter abbreviation: D (demand/load), L (line), G (generator), or T (transformer).  The model requires 2.44 seconds to solve, including pre-processing.  The objective function value of the output solution is -3805.  If each timestep represents a real time of one minute, this value represents the total energy served over the restoration horizon in MW-min.

\newcolumntype{C}[1]{>{\centering\arraybackslash}p{#1}}
\begin{table*}[ht]
  \centering
  \caption{Results of OPF-R on IEEE 9-Bus Model.}
  \footnotesize
  \resizebox{\columnwidth}{!}{
  \begin{tabular}{c|c|c|c|c|c|c|c|c|c|c|c|c|c|c|c|c|c|c|c|c|c}
    \toprule
    Timestep & Init. & 2 & 3 & 4 & 5 & 6 & 7 & 8 & 9 & 10 & 11 & 12 & 13 & 14 & 15 & 16 & 17 & 18 & 19 & 20 & 21 \\ \midrule
    Component & G & T & L & D & D & D & L & D & D & L & T & G & D & L & D & D & D & L & L & - & - \\ \midrule
    Buses & 1 & 1--4 & 4--5 & 5 & 5 & 5 & 4--6 & 6 & 6 & 5--7 & 7--2 & 2 & 6 & 7--8 & 8 & 8 & 8 & 6--9 & 9--8 & - & - \\ \bottomrule
  \end{tabular}
  }
  \label{tab:9bus-static}
\end{table*}

\begin{table*}[ht]
  \centering
  \caption{Results of OPF-R on IEEE 9-Bus Model, 2 loads per bus, 10 timesteps.}
  \footnotesize
  \begin{tabular}{c|c|c|c|c|c|c|c|c|c|c}
    \toprule
    Timestep & 1 & 2 & 3 & 4 & 5 & 6 & 7 & 8 & 9 & 10  \\ \midrule
    Component & T & L & D & D & L & D & L & T & G & D \\ \midrule
    Buses & 1--4 & 4--5 & 5 & 5 & 4--6 & 6 & 6--9 & 9--3 & 3 & 6 \\ \bottomrule
  \end{tabular}
  \label{tab:9bus-static-2load}
\end{table*}

It is observed from the output switching order that the formulation performs as expected.  First, the three largest loads of 41.67 MW each are energized at Bus 5.  Then, two loads of 30 MW are picked up at Bus 6, before Generator 1 reaches its limit of 200 MW.  Generator 2 is added to increase production capacity, followed by pickups of the remaining loads.

It is notable that more than half of the loads, three at Bus 5 and two at Bus 6, are switched on with only one generator energized.  In realistic restoration operations, this configuration is unlikely to be dynamically feasible due to the low frequency nadir that is expected when only one synchronous generator is connected.  However, this result is coherent with the problem formulation: the optimization is incentivized to energize loads as early as feasible, given static active power constraints.  This aspect is further elaborated in the next section.

It is noted that not all of the 20 components in the model are energized within the restoration horizon, despite there being enough time allocated to the model to do so.  This is an artifact of the objective function: after all loads are energized, there is no incentive to energize additional components.  Thus, that the global optimum for this problem is non-unique, as the energization order after timestep 17 has no impact, and further, there is no advantage to picking up Generator 3 over Generator 2, as the same number of switches is required.

As the presented formulation is non-convex, it is prudent to validate the global optimality of the switching constraints.  This is accomplished by generating possible combinations of the switching variables $S_d$, $S_g$, $S_l$, and $S_w$, then solving a Linear Program (LP) where the switching variables are parameters.  Specifically, this validation LP consists of Constraints~(\ref{c:powerbal}), (\ref{c:branchp}), and the component limits.  Given that the validation LP is a convex problem, convergence to a global optimum is guaranteed~\cite{conv-opt}.  The set of solutions with feasible validation LPs is then compared in terms of the value of the given objective~(\ref{eq:obj}) to find the globally optimal switching sequence.

Although it is methodologically simpler to generate the switching variable options by matrix manipulation, this results in an exponentially higher number of validation LPs to solve.  Specifically, the network in question has 20 components.  Assuming that the entire network is to be energized, the number of options without considering connectivity is $20! = 2.43 \textrm{E} 18$.  Therefore, the in-order transversal (depth first) tree search method~\cite{Morris1979} is implemented to limit the options to those which are feasible given network connectivity, but disregarding power flow constraints.  The order of the load switches at a given bus is disregarded, as their magnitudes are equal.

Still, to ensure reasonable computational time, the validation case is selected as the version of the model with 2 loads per bus (i.e., 62.5 MW, 45 MW, and 50 MW at Buses 5, 6, and 8, respectively).  The first 10 timesteps of the restoration process are optimized.  The results of the optimization are shown in Table~\ref{tab:9bus-static-2load}, where the switching order is observed to be similar as for the case with 3 loads per bus.  This case requires 0.79 seconds to solve, including pre-processing, and the final objective function value is -1207.5.

Without considering connectivity, the number of possible options is $17!/(17-10)! = 7.06\textrm{E}10$. Considering connectivity, the number of possible options is 240,800.  Of these, 183,317 are determined to be feasible using the described validation LP method.  The global optimum has an objective function value of -1207.5, and includes the same load energization order as the solution in Table~\ref{tab:9bus-static-2load}, validating the proposed method.  However, Generator 2 is energized at timestep 9 instead of Generator 3, indicating a non-unique global optimum, as mentioned.

\section{OPF for Restoration Considering Frequency Dynamics: DynOPF-R}\label{chap:dynopf-r}

As anticipated, static formulations of the optimal restoration problem may result in solutions which are operationally infeasible due to frequency transients.  Specifically, it is supposed that the static optimal restoration path for the IEEE 9-Bus model, in which 5 loads are energized before a second generator, would result in trips due to underfrequency protection.  For this purpose, a dynamic system model is developed, to test this hypothesis, as well as to compute the optimal restoration path considering power system dynamics.  

In this work, the dynamic model is inspired by standard governor models of synchronous machines, as has been demonstrated in the literature for applications to security-constrained OPF problems~\cite{Zhao_Bialek2016,Alvarez2024}.  The implemented dynamic model is shown in Fig.~\ref{fig:dyn_blockdiag}.  Notably, it is reasonable to assume that secondary control is deactivated during the PSR process~\cite{CH-TC}.

% Adapted from: 
% https://tex.stackexchange.com/questions/175969/block-diagrams-using-tikz
% https://tex.stackexchange.com/questions/494953/tikz-block-diagram-with-summation-block-having-crossed-lines
\tikzset{
block/.style = {draw, fill=white, rectangle, minimum height=3em, minimum width=3em},
tmp/.style  = {coordinate}, 
sumA/.style = {circle, draw, minimum size=9mm,
        append after command={\pgfextra{\let\LN\tikzlastnode}
            (\LN.north west) edge (\LN.south east)
            (\LN.south west) edge (\LN.north east)
        node[left]     at (\LN.center) {$+$}
        node[below]    at (\LN.center) {$-$}
                                    }},
sumB/.style = {circle, draw, minimum size=9mm,
        append after command={\pgfextra{\let\LN\tikzlastnode}
            (\LN.north west) edge (\LN.south east)
            (\LN.south west) edge (\LN.north east)
        node[left]     at (\LN.center) {$-$}
        node[below]    at (\LN.center) {$-$}
        node[above]    at (\LN.center) {$+$}
                                    }},
input/.style = {coordinate},
output/.style= {coordinate},
pinstyle/.style = {pin edge={to-,thin,black}}
}
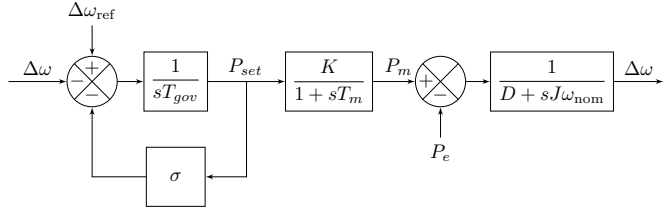
\begin{figure}[t]
\centering
\resizebox{0.5\columnwidth}{!}{
\begin{tikzpicture}[auto, node distance=2cm,>=latex']
    \node [input, name=rinput] (rinput) {};
    \node [sumB, right of=rinput, node distance=1.2cm] (sum1) {};
    \node [block, right of=sum1, node distance=1.5cm] (int) {$\dfrac{1}{sT_{gov}}$};
    \node [block, right of=int,node distance=2.75cm] (turb) {$\dfrac{K}{1+sT_m}$};
    \node [block, below of=int,node distance=1.7cm] (sigma) {$\sigma$};
    \node [block, below of=sigma,node distance=1.7cm] (delta) {$\dfrac{\delta sT_r}{1+sT_r}$}; 
    \node [sumA, right of=turb,node distance=2cm] (sum3) {};
    \node [block, right of=sum3,node distance=2cm] (swing) {$\dfrac{1}{D+sJ \omega_\textrm{nom}}$};
    \node [output, right of=swing, node distance=1.9cm] (output) {};
    \draw (4,0) node[above] {$P_{set}$};
    \draw (6.6,0) node[above] {$P_{m}$};
    \draw [->] (rinput) -- node{$\Delta \omega$} (sum1);
    \draw [->] (sum1) -- (int);
    \draw [->] (int) --node[name=z,anchor=south]{} (turb);
    \draw [->] (turb) -- (sum3);
    \draw [->] (sum3) -- (swing);
    \draw [->] (swing) -- node [name=y] {$\Delta \omega$}(output);
    \draw [->] (z) |- (sigma);
    \draw [->] (sigma) -| (sum1);
    \draw [->] (z) |- (delta);  
    \draw [->] (delta) -| (sum1);  
    \draw [->] ($(sum3)-(0,1.0cm)$)node[below]{$P_e$} -- (sum3);
    \draw [->] ($(0,1.0cm)+(sum1)$)node[above]{$\Delta \omega_{\mathrm{ref}}$} -- (sum1);
    \end{tikzpicture}
    }
\caption{Block diagram of the implemented governor, turbine, and generator dynamics.} \label{fig:dyn_blockdiag}
\end{figure}

As depicted, the dynamic constraints which define the system frequency consist of three components: the governor model, the Synchronous Machine (SM) model, and the swing equation, which represents the mechanical dynamics of the SM and prime mover.  As each SM may have its own mechanical rotational speed during PSR operations, the frequency is modelled at each defined generator bus (i.e., PU and slack buses), with bus frequencies expected to converge to similar values. The swing equation is implemented as follows:
\begin{equation}\label{eq:swing} 
    M_g \dot{\Delta \omega}_g = P_{m,g} - P_{e,g} - \hat{D}_g \cdot (\Delta \omega_g + \omega_\textrm{nom}) \quad \forall g \in [\mathcal{B},\mathcal{N}]
\end{equation}

\noindent where $\Delta \omega_g$ is the frequency at generator bus $g$ relative to the synchronous frequency in \si{rad/s}, $\omega_\textrm{nom}$ is the synchronous frequency of the system, $M_g$ is the angular momentum of machine $g$, and $\hat{D}_g$ is the absolute speed damping constant of machine $g$ in terms of power.  Added variables $P_m$ and $P_e$ denote the mechanical and electrical power of the machine.

To couple the dynamic equations to the DC OPF constraints, some minor modifications are required.  First, the power balance Constraint~(\ref{c:powerbal}) is modified as follows (adapted from~\cite{Zhao_Bialek2016}):
\begin{equation}\label{eq:dyn-powerbal}
    \mathbf{P}_{e} = \mathbf{I}_d (\mathbf{S}_d \circ \mathbf{P}_d) + \mathbf{I}_l \mathbf{P}_{l} + \mathbf{I}_w \mathbf{P}_{w}
\end{equation}

\noindent where $P_{e,b,t}$ defines the electrical power balance at bus $b$ and time $t$, which is represented in matrix form as $\mathbf{P}_e$.  For load buses, it is specified that $P_e = 0$.  To link the bus power angles $\theta$ with the generator dynamic equations, the relation between the power angle and frequency is defined as follows:
\begin{equation}\label{eq:delta}
    \dot{\theta_g} = \omega_\textrm{nom} + \Delta \omega_g \quad \forall g \in [\mathcal{B},\mathcal{N}]
\end{equation}

Eq.~(\ref{eq:delta}) also ensures convergence of the generator bus frequencies, by construction.  Further, the mechanical power balance of each generator is modelled using the following first-order differential equation:
\begin{equation}\label{eq:gen}
    \dot{P}_{m,g} = \frac{1}{T_{m,g}} (K_g P_{set,g} - P_{m,g}) \quad \forall g \in [\mathcal{B},\mathcal{N}]
\end{equation}

\noindent where $K_g$ is the gain and $T_{m,g}$ is the time constant of generator $g$\footnote{The time constant of the generator $T_m$ includes the SM, prime mover, and coupling shaft.}.  The turbine governor includes a transient droop component, which can be implicitly included in optimization problem constraints by introducing a variable $Y$ representing the signal at the output of the transient droop block.  The associated equations are derived as follows:
\begin{equation}\label{eq:gov}
\begin{split}
    T_{gov,g} \dot{P}_{set,g} &= \Delta \omega_{\textrm{ref},g} - \Delta \omega_g - \sigma_g P_{set,g} - Y_g  \\ &\quad \forall g \in [\mathcal{B},\mathcal{N}] 
    \end{split}
\end{equation}
\begin{equation}\label{eq:Yg-orig}
    T_{r,g} \dot{Y}_g = \delta_g T_{r,g} \dot{P}_{set,g} - Y_g \quad \forall g \in [\mathcal{B},\mathcal{N}] 
\end{equation}

\noindent where $T_{gov}$ is the governor time constant,  $\sigma$ is the steady-state droop, $T_r$ is the transient droop time constant, and $\delta$ is the transient droop gain.  The setpoint $\Delta \omega_\textrm{ref}$ is defined for each generator and is a decision variable.  As the transient droop component decays to zero in steady-state, the frequency should settle to a value which is proportional to $-1/\sigma$. 

It can be anticipated that the computation of the numerical derivative of $P_{set}$ will negatively impact the numerical properties of the formulation.  To alleviate this, Eq.~(\ref{eq:Yg-orig}) is replaced by the following, derived by combining Eq.~(\ref{eq:gov}) and~(\ref{eq:Yg-orig}):
\begin{equation}\label{eq:Yg-v2}
    \begin{split}
    \dot{Y}_g = &\frac{\delta_g}{T_{gov,g}} (\Delta \omega_{\textrm{ref},g} - \Delta \omega_g)
    - \frac{\delta_g \sigma_g}{T_{gov,g}} P_{set,g} \\
    & - \left(\frac{\delta_g}{T_{gov,g}} + \frac{1}{T_{r,g}}\right) Y_g
    \quad \forall g \in [\mathcal{B},\mathcal{N}] 
    \end{split}
\end{equation}

Finally, the dynamic variables are subject to physical limits:
\begin{equation}\label{eq:pg_lim}
    \underline{P_{g}} \leq P_{e,g} \leq \overline{P_{g}} \quad \forall g \in [\mathcal{B},\mathcal{N}], t \in \mathcal{T}
\end{equation}
\begin{equation}\label{eq:flim}
    2 \pi \Delta f_{min} \leq \Delta \omega_g \leq 2 \pi \Delta f_{max} \quad \forall g \in [\mathcal{B},\mathcal{N}], t \in \mathcal{T}
\end{equation}

\noindent where $\Delta f_{max}$ is the maximum frequency deviation from nominal, chosen by the modeler as an optimization parameter.

The practical implementation of these constraints requires the use of numerical integration.  To keep the constraints linear and maximize numerical stability, the Backward Euler method is deployed for Eq.~(\ref{eq:swing}), (\ref{eq:gen}), (\ref{eq:gov}), and (\ref{eq:Yg-v2}), and the constraints are written in per unit. The discretized formulation is as follows:
\begin{equation}\label{eq:swing_imp}
    \begin{split}
    \Delta \omega_{g,t} &= \Delta \omega_{g,t-1} + \frac{\Delta t}{M_g} ( P_{m,g,t} - P_{e,g,t} \\
     &  - \hat{D}_g \cdot (\Delta \omega_{g,t} + \omega_\textrm{nom})) \quad \forall g \in [\mathcal{B}, \mathcal{N}], t \in \mathcal{T}'
    \end{split}
\end{equation}
\begin{equation}
    \begin{split}
    P_{m,g,t} = & P_{m,g,t-1} + \frac{\Delta t}{T_{m,g}} (K_g P_{set,g,t} - P_{m,g,t})  \\ & \quad \forall g \in [\mathcal{B}, \mathcal{N}], t \in \mathcal{T}'
    \end{split}
\end{equation}
\begin{equation}
    \begin{split}
    P_{set,g,t} = &{P}_{set,g,t-1} - \frac{\Delta t}{T_{gov,g}} ( \Delta \omega_{g,t} - \Delta \omega_{\textrm{ref},g}  \\
    & + \sigma_g P_{set,g,t} + Y_{g,t})  \quad \forall g \in \mathcal{B}, \mathcal{N}, t \in \mathcal{T}'
    \end{split} 
\end{equation}
\begin{equation}
    \begin{split}
    Y_{g,t}=& Y_{g,t-1} + \frac{\Delta t \cdot \delta_g}{T_{gov,g}} (\Delta \omega_{\textrm{ref},g} - \Delta \omega_{g,t}) \\
    & - \frac{\Delta t \cdot \delta_g \sigma_g}{T_{gov,g}} P_{set,g,t} - \left(\frac{\Delta t \cdot \delta_g}{T_{gov,g}} + \frac{\Delta t}{T_{r,g}}\right) Y_{g,t} \\
    & \forall g \in [\mathcal{B},\mathcal{N}], t \in \mathcal{T}'
    \end{split}
\end{equation}

Furthermore, big-M constraints are applied to the model to handle the initialization of the generators and computation of the power angles (Eq.~(\ref{eq:delta})).  Big-M constraints rely on a parameter M, which is set to a large value (e.g., 1E6), and a binary variable which controls whether, or not, the constraint is active.  The constraints specified by Eq.~(\ref{eq:sw-gen-bigm}),~(\ref{eq:theta-bsu})--(\ref{eq:theta-nbsu2}) are required to ensure feasibility of the optimization problem at the moment of generator pickup. Similar methods have been deployed for constraining the synchronization of two independent microgrids~\cite{Maharjan2025}.

The following constraint ensures that, at the moment of generator energization, the frequency of the generator matches the frequency of the system.
\begin{equation}\label{eq:sw-gen-bigm}
    \begin{split}
    -M(1-&S_{j,t}-S_{j,t-1}) \leq \omega_{i,t}-\omega_{j,t} 
    \\ & \leq M(1-S_{j,t}-S_{j,t-1}) \quad \forall j \in \mathcal{N}, t \in \mathcal{T}'
    \end{split}
\end{equation}

\noindent where $i$ is the index of the leading black-start generator.  It is noted that the principle behind the constraint is representative of the reality, as NBSUs must be ramped to the synchronous frequency of the system before being switched in to the island, commonly via synchrocheck switching relay.  The following constraint enforces that the electrical power balance is zero at generator buses before the generators are energized:
\begin{equation}\label{eq:pbal-gen-bigm}
    -M S_{g,t} \leq P_{e,g,t} \leq M S_{g,t} \quad \forall g \in [\mathcal{B},\mathcal{N}], t \in \mathcal{T}
\end{equation}

The power angle constraints are implemented such that a bus angle is only coupled to its frequency if the relevant NBSU is switched on, as follows:
\begin{equation}\label{eq:theta-bsu}
    \theta_{i,t} - \theta_{i,t-1} = \Delta t (\omega_\textrm{nom} + \Delta \omega_{i,t}) \quad \forall i \in \mathcal{B}, t \in \mathcal{T}'
\end{equation}
\begin{equation}\label{eq:theta-nbsu1}
    \begin{split}
    \frac{\theta_{j,t} - \theta_{j,t-1}}{\Delta t} &- (\omega_\textrm{nom} + \Delta \omega_{j,t}) - \epsilon \\
    & \leq M(1-S_{j,t}) \quad \forall j \in \mathcal{N}, t \in \mathcal{T}'
    \end{split}
\end{equation}
\begin{equation}\label{eq:theta-nbsu2}
    \begin{split}
    \frac{\theta_{j,t} - \theta_{j,t-1}}{\Delta t} &- (\omega_\textrm{nom} + \Delta \omega_{j,t}) + \epsilon \\
    & \geq -M(1-S_{j,t}) \quad \forall j \in \mathcal{N}, t \in \mathcal{T}'
    \end{split}
\end{equation}

\noindent where $\epsilon$ is a tolerance which can be used to relax the constraints, as applying a strict (zero) convergence threshold to the generator frequencies can cause the optimization problem to be infeasible.

As the implementation is based on discretized differential equations, it is important to initialize the state variables $P_{set}$, $P_m$, $Y$, and $\Delta \omega$.  The initialization is implemented as follows:
\begin{equation}
    \theta_{i,0} = 0
\end{equation}
\begin{equation}
    Y_{g,0} = 0 \quad \forall g \in [\mathcal{B}, \mathcal{N}] 
\end{equation}
\begin{equation}
    \Delta \omega_{g,0} = \frac{P_{m,g,0}-P_{e,g,0}}{\hat{D}_g}-1 \quad \forall g \in [\mathcal{B}, \mathcal{N}] 
\end{equation}
\begin{equation}
    P_{set,g,0} = \frac{\Delta \omega_{\textrm{ref},g}-\Delta \omega_{g,0}}{\sigma_g} \quad \forall g \in [\mathcal{B}, \mathcal{N}] 
\end{equation}
\begin{equation}\label{eq:pm_init}
    P_{m,g,0} = K_g P_{set,g,0} \quad \forall g \in [\mathcal{B}, \mathcal{N}]
\end{equation}

\noindent where $i$ is the index of the leading black-start generator. In contrast with the static formulation, the power angle of the slack bus is constrained to zero only at $t=0$, as the power angle must be permitted to spin with angular frequency $\omega$.

In the dynamic version of the problem, the timesteps and time limit for the restoration process are implemented in absolute terms (seconds).  Therefore, when considering the restoration process from an operational point of view, it makes sense to consider a certain minimum amount of time which should be required between switches (``dead time''), defined as a parameter $\kappa$ (seconds) which can be tuned by the modeler.  

This enables a dimensionality reduction of the switching constraints, which significantly improves the computational efficiency of the branch-and-bound algorithm, implemented as follows.  First, a set $\mathcal{T}^*$ is defined to represent the set of switching timesteps, such that $\mathcal{T}^* = [1,2,...,n_r] \in \mathcal{T}$, where the length of $\mathcal{T}^*$ is $n_r = \lfloor T_{max} \Delta t / \kappa \rfloor + 1$.  Condensed switching variables are introduced as $\mathbf{R}_d, \mathbf{R}_g, \mathbf{R}_l, \mathbf{R}_w$ for loads, generators, lines, and transformers, respectively, where the dimension of a matrix $\mathbf{R}$ is $n_i \times n_r$.  Then, Constraints~(\ref{c:0init})--(\ref{c:bus_lines}) are modified such that they are defined on domain $\mathcal{T}^*$, and a constraint is added which maps the relation between variables $\mathbf{S}$ and $\mathbf{R}$, as follows:
\begin{equation}\label{eq:mapping}
    S_{i,t} = R_{i,\lfloor t \Delta t/\kappa + 1 \rfloor} \quad \forall i \in [\mathcal{B}, \mathcal{N}, \mathcal{D}, \mathcal{L}, \mathcal{W}], t \in \mathcal{T}
\end{equation}

Finally, although Eq.~(\ref{eq:flim}) bounds the frequency nadir, it is technically preferable to apply a penalty on the frequency deviations in addition to the bonus for energy served.  The objective function is modified as follows:
\begin{equation}\label{eq:obj-dyn}
    \alpha \Delta t \sum_{d,t} S_{d,t} P_d - \beta \Delta t \sum_{g,t} |\Delta \omega_{g,t}| 
\end{equation}

\noindent where $\alpha$ and $\beta$ are objective function weights which must be tuned by the modeler.  It is noted that all of the added constraints, with the exception of Eq.~(\ref{eq:dyn-powerbal}), are linear.  The power balance,  Eq.~(\ref{eq:dyn-powerbal}), and the objective function, Eq.~(\ref{eq:obj-dyn}), are mixed-integer bilinear.

In summary, the formulation of DynOPF-R is as follows:
\begin{subequations}
    \begin{alignat}{4}
        &\max_{S,R,P_e,P_m,P_{set},P_l,P_w,\theta,\omega} & \quad & \textrm{Eq. (\ref{eq:obj-dyn})} \\
        &\text{s.t.} & &  \textrm{Eq.~(\ref{eq:dyn-powerbal}), (\ref{eq:pg_lim})--(\ref{eq:mapping})} \\ 
        & & &  \textrm{Constr.~(\ref{c:branchp})--(\ref{c:bus_lines}), (\ref{c:linelims})--(\ref{c:trafolims})} 
    \end{alignat}
\end{subequations}

\section{Application of DynOPF-R on the IEEE 9-Bus Model}\label{chap:results}

In this Section, the proposed DynOPF-R formulation is implemented on the IEEE 9-Bus model, as described and pictured in Section~\ref{chap:static}.  For this use case, the choice of a small, transmission-level system is reasonable to represent the early stages of the restoration process (i.e., the creation of restoration islands); a known example is used for reproducibility. 

\subsection{Parameters' Inference for the Dynamic OPF Constraints} \label{chap:inference}

The dynamic model, which forms the constraints of the optimization problem, is pictured in Fig.~\ref{fig:dyn_blockdiag}.  It can be seen that the dynamic parameters $D$, $J$, $K$, $T_m$, $T_{gov}$, $T_r$, $\delta$, and $\sigma$ must be selected, where possible, to match the parameters of realistic synchronous machines. To this end, an adaptation of the IEEE 9-Bus dynamic model~\cite{psanalysis} is used to infer the optimization problem parameters, which has been made available at~\cite{github-9bus}.  

The reference EMT model~\cite{github-9bus} includes an IEEEG3 turbine governor model, for which the parameters are slightly adapted for these experiments.  The turbine governor model is depicted in Fig.~\ref{fig:gov}, and the parameters for the inference experiments are shown in Table~\ref{tab:params-emtp} for each of the three generators $SM_1$, $SM_2$, and $SM_3$.  Time constants (TC) are in units of seconds, while gains and coefficients (CF) are in per unit.  The inertia constant $H$ and damping coefficient D are SM parameters and are not depicted in the control scheme.  

\begin{table}[t]
\caption{Turbine-governor parameters for the reference EMT model.}
\centering
\renewcommand{\arraystretch}{1.1}
\begin{tabular}{C{3.75cm} C{0.75cm} C{0.75cm} C{0.75cm}}
\toprule
Parameter & $SM_1$ & $SM_2$ & $SM_3$ \\ \midrule
Inertia constant $H$ [s]            & 4.5    & 3.5    & 6    \\   
Speed damping $D$ [Nms/rad] & 6780 & 6780  & 6780 \\   
Gate gain $K_g$ [p.u.]              & 5    & 5   & 5    \\   
Valve TC $T_p$ [s]                  & 0.05 & 0.05 & 0.05  \\   
Transient droop TC $T_r$ [s]        & 5    & 5    & 5    \\    
Permanent droop CF $K_\sigma$ [p.u.]  & 0.02 & 0.03 & 0.04  \\ 
Transient droop CF $K_\delta$ [p.u.]  & 0.8  & 0.8  & 0.8   \\ 
Turbine gain $K_t$ [p.u.]           & 2    & 1.75  & 1.5  \\ 
Turbine numerator TC $T_n$ [s]      & -0.2    & -0.2    & -0.2    \\ 
Turbine denominator TC $T_d$ [s]    & 2 & 1.5 & 2.4  \\ \bottomrule
\end{tabular}
\label{tab:params-emtp}
\end{table}

\tikzset{
block/.style = {draw, fill=white, rectangle, minimum height=3em, minimum width=3em},
tmp/.style  = {coordinate}, 
sum/.style= {draw, fill=white, circle, node distance=1cm},
input/.style = {coordinate},
output/.style= {coordinate},
pinstyle/.style = {pin edge={to-,thin,black}}
}
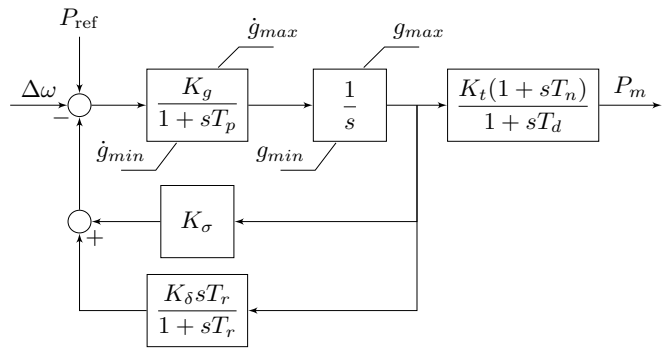
\begin{figure}[t]
\centering
\resizebox{!}{4cm}{
\begin{tikzpicture}[auto, node distance=2cm,>=latex']
    \node [input, name=rinput] (rinput) {};
    \node [sum, right of=rinput] (sum1) {};
    \node [block, right of=sum1, node distance=1.7cm] (servo) {$\dfrac{K_g}{1+sT_p}$};
    \node [block, right of=sum1, node distance=3.9cm] (int) {$\dfrac{1}{s}$};
    \node [block, right of=int,node distance=2.5cm] (turb) {$\dfrac{K_t (1+sT_n)}{1+sT_d}$};
    \node [block, below of=servo,node distance=1.7cm] (sigma) {$K_\sigma$};
    \node [block, below of=sigma,node distance=1.3cm] (delta) {$\dfrac{K_\delta sT_r}{1+sT_r}$};
    \node [sum, below of=sum1,node distance=1.7cm] (sum2) {};
    \node [output, right of=turb, node distance=2cm] (output) {};
    \draw [->] (rinput) -- node{$\Delta \omega$} (sum1);
    \draw [->] (sum1) -- (servo);
    \draw [->] (servo) -- (int);
    \draw [->] (int) --node[name=z,anchor=south]{} (turb);
    \draw [->] (sigma) -- node[pos=0.95] {$+$} (sum2);
    \draw [->] (turb) -- node [name=y] {$P_m$}(output);
    \draw [->] (z) |- (delta);
    \draw [->] (z) |- (sigma);
    \draw [->] (delta) -| (sum2);
    \draw [->] (sum2) -- node[pos=0.99] {$-$} (sum1);
    \draw [->] ($(0,1.0cm)+(sum1)$)node[above]{$P_{\mathrm{ref}}$} -- (sum1);
    \draw (3,0.55) -- (3.3,0.85) -- (3.8,0.85) node[above]{$\dot{g}_{max}$};
    \draw (5.1,0.55) -- (5.4,0.85) -- (5.9,0.85) node[above]{${g}_{max}$};
    \draw (2.4,-0.55) -- (2.1,-0.95) -- (1.6,-0.95) node[above]{$\dot{g}_{min}$};
    \draw (4.7,-0.55) -- (4.4,-0.95) -- (3.9,-0.95) node[above]{${g}_{min}$};
    \end{tikzpicture}
    }
\caption{Block diagram of implemented governor/turbine model IEEEG3~\cite{turb-gov-argonne}.} \label{fig:gov}
\end{figure}

Some governor parameters have been changed from the original version in~\cite{github-9bus}, outlined as follows.  First, it is interesting to consider generators with different dynamic responses with respect to each other for the optimization problem; for this reason, the parameters $H$, $K_\sigma$, $K_t$, and $T_d$ are adapted in the reference EMT model.  The parameter $T_n$ is then reduced in absolute value to achieve dynamic stability in a reasonable timeframe.  Finally, the MVA rating of all generators is increased to 200 MVA. All other model parameters are unchanged from the original version.  

\begin{figure}[h!]
    \centering
    \subfloat[]{\includegraphics[width=8.8cm]{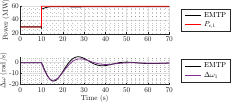}}
    \label{subfig:1}
    \subfloat[]{\includegraphics[width=8cm]{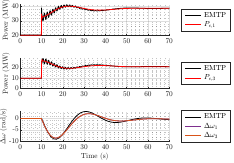}}
    \label{subfig:2}
    \subfloat[]{\includegraphics[width=8cm]{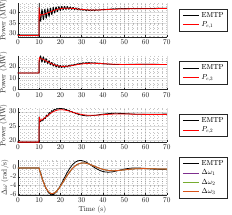}}
    \label{subfig:3}
    \caption{Results of the inference experiments: (a) Experiment 1 (b) Experiment 2 (c) Experiment 3.}
    \label{fig:emtp-valid}
\end{figure}

\begin{table}[t]
\caption{Dynamic Parameters of Generators, DynOPF-R.}
\centering
\renewcommand{\arraystretch}{1.1}
\begin{tabular}{C{1cm} C{1.8cm} C{1.15cm} C{1.15cm} C{1.15cm}}
\toprule
Param. & Unit & Value $G1$ & Value $G2$ & Value $G3$ \\ \midrule
$M$  & \si{MW-s^2/rad}   & 5.7296 & 4.4563 & 7.6394    \\ 
$\hat{D}$  & \si{MW-s/rad}     &  0.426 & 0.426 & 0.426    \\ 
$T_g$  & \si{s}   & 2 & 1.5 & 2.4     \\ 
$K$  & p.u.   & 2 & 1.75 & 1.5     \\ 
$T_{gov}$   & \si{s}    & 0.6 & 0.8 & 0.9     \\ 
$T_r$   & \si{s}    & 5 & 5 & 5     \\ 
$\delta$   & p.u.    & 0.8 & 0.8 & 0.8     \\ 
$\sigma$   & p.u.    & 0.02 & 0.03 & 0.04      \\ \bottomrule
\end{tabular}
\label{tab:dyn-param}
\end{table}

All three experiments include one load of 30 MW initially energized at Bus 6, while a second load of 30 MW is energized at $t=10$ seconds.  Experiment 3 includes an additional load of 33.33 MW initially energized at Bus 8.  

The dynamics produced by the optimization problem are evaluated by providing the switching variables to the formulation as parameters, excluding the generation and branch flow limits.  As generator switching is out of scope, the big-M constraints (i.e., Eq.~(\ref{eq:sw-gen-bigm}), (\ref{eq:pbal-gen-bigm}),~(\ref{eq:theta-nbsu1}), and~(\ref{eq:theta-nbsu2})) are not needed.  The power angles are treated together using Eq.~(\ref{eq:theta-bsu}).  A discretization time $\Delta t$ of 100 ms is used for the inference process, to minimize discretization error, yet remain above the timescale of electromagnetic transients~\cite{sauer-pai}.

For the inference process, initial electrical power generation $P_{e,0,2}$ and $P_{e,0,3}$ are fixed as parameters of the optimization problem.  For Experiment 2, $P_{e,0,3} = 10$, and for Experiment 3, $P_{e,0,3} = 14.61$ and $P_{e,0,2} = 19.49$.  These values are derived from the steady-state droop of the energized machines relative to the initially energized loads, and are replicated in the EMTP load flow solution used to initialize the time domain simulation.  It is also imposed that $\Delta \omega_{1,t=0} = 0$.

Most dynamic model parameters can be extracted from the EMTP model directly.  The angular momentum of generators $M$ is computed according to the following formula:
\begin{equation}
    M_g = \frac{2H\overline{P_g}}{\omega_\textrm{nom}} \quad \forall g \in [\mathcal{B},\mathcal{N}]
\end{equation}

Further, the absolute speed damping constant in terms of power $\hat{D}$ is computed according to the following formula:
\begin{equation}
    \hat{D}_g = \frac{2 D_g \omega_\textrm{nom}}{p_g} \quad \forall g \in [\mathcal{B},\mathcal{N}]
\end{equation}

\noindent where $p_g$ is the generator's number of poles.  The parameter $T_{gov}$ is then inferred from the measured data.  The results of the described inference process are shown in Fig.~\ref{fig:emtp-valid}, and the inferred optimization parameters are shown in Table~\ref{tab:dyn-param}.  The rotational speed and electrical power of each machine in EMTP is extracted using meters which are built-in to EMTP's synchronous machine model~\cite{EMTP-SM}.

It is observed that the initialization equations of the model function as expected: the frequency and active power remain stable until the load step is applied.  The frequency and active power transients produced by the model well-approximate its EMT counterparts.  Notable sources of error in the optimization problem relative to the reference EMT model are the DC approximation (reactive power and voltage dynamics are neglected), as well as the simplification of the gate model to first-order.  Thus, while this analysis shows that the formulation can well-approximate the  frequency dynamics of a detailed EMT model with AC components, the mentioned sources of error necessitate the described parameter inference process to achieve the former.  It is further noted that the DC approximation may not be valid for distribution networks due to higher R/X ratios. 

\subsection{Application of DynOPF-R to the 9-Bus Model}

In this section, the dynamic optimization problem, as described in Section~\ref{chap:dynopf-r}, is run on the IEEE 9-Bus model using the inferred generator parameters from Section~\ref{chap:inference} (Table~\ref{tab:dyn-param}), and the results are analyzed.

\begin{figure}[t]
    \centering
    \includegraphics[width=8.8cm]{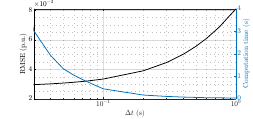}
    \caption{Results of DVLP with load energizations at Bus 5.}
    \label{fig:dt-analysis}
\end{figure}

As the frequency and generator dynamics are modelled by discretized differential equations, it can be shown that there is a trade-off between the accuracy of the frequency estimation (i.e., the magnitude of the discretization error) and the time required for computation of the dynamic state variables.  To illustrate this, we simulate a single load pickup with varying $\Delta t$ and observe the frequency transient in comparison with the ground truth response, also shown in Fig.~\ref{fig:emtp-valid}.  The error of the optimized frequency transient relative to the EMTP simulation is quantified using root mean squared error (RMSE), computed on the range $t = [t_{switch}, t_{switch} + 60s]$.  The results of this analysis are shown in Fig.~\ref{fig:dt-analysis}.

It is observed that the lower the $\Delta t$, the closer the frequency response to the ground truth.  Therefore, it is foreseeable that the discretization error could cause the branch and bound algorithm to reach a bifurcation point and therefore converge to a different optimal switching order.  Thus, to ensure computational efficiency, the maximum $\Delta t$ which satisfies the specified accuracy range should be chosen.

For this Case Study, the parameters of the optimization problem are specified in Table~\ref{tab:opt-param}.  The chosen time discretization $\Delta t$ is 0.2 seconds, such that a reasonable trade-off between discretization error and computational time is achieved; at this point, the RMSE of the frequency response is 0.11\% higher than the RMSE at discretization error convergence (evaluated at $\Delta t = 5 ms$). The number of timesteps in the model is 3600, such that the first 15 switches are optimized.  The frequency tolerances are  $\Delta f_{max} = 1.5 \si{Hz}$ and $\Delta f_{min} = -2 \si{Hz}$, thus, the permitted frequency range is $[48, 51.5]$ $\si{Hz}$. The overfrequency limit of 51.5 Hz corresponds to the automatic disconnection point of generators~\cite{CH-TC}, while the underfrequency limit of 48 Hz is chosen to lie at the lower limit of the range for underfrequency load shedding~\cite{CH-ufls}.  In practice, the modeler should choose the active power and frequency limits for their system such that unexpected trips are avoided.  The simulation is carried on a desktop computer with a 3.2 GHz Intel processor and 64 GB of RAM.

\begin{table}[ht]
\caption{Case Study Optimization Parameters.}
\centering
\renewcommand{\arraystretch}{1.1}
\begin{tabular}{C{3cm} C{1cm} C{1cm}}
\toprule
Parameter & Value & Unit \\ \midrule
$\Delta f_{max}$  &  1.5  & \si{Hz}    \\ 
$\Delta f_{min}$  &  -2  & \si{Hz}    \\ 
$T_{max}$   &  12   & \si{min}     \\ 
$\Delta t$   &  0.2   & \si{s}      \\ 
M (big-M constraints)   &   1E6  & -      \\ 
$\alpha$,  $\beta$   &  1   & -      \\ 
$\epsilon$   &  0.05   & \si{rad/s}      \\
$\kappa$   & 45    & \si{s}      \\ \bottomrule
\end{tabular}
\label{tab:opt-param}
\end{table}

\begin{figure*}[t]
    \centering
    \resizebox{16cm}{7.45cm}{
    \begin{tikzpicture}[font=\scriptsize]
        \node[anchor=south west, inner sep = 0] at (0,0) {\includegraphics[width=16cm]{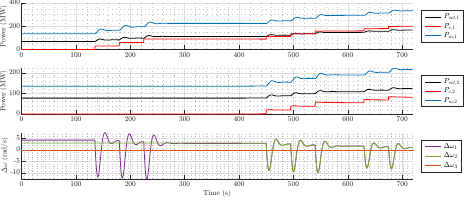}};
        %\node[anchor=south west, inner sep = 0] at (0,0) {\includegraphics[width=16cm]{fig/dyn_result_opf_final.pdf}};

        % Define coordinates
        \coordinate (START) at (0.74,-0.2);
        \coordinate (700) at (13.8,-0.3); % used to compute the coords
        \coordinate (sw1) at ($(START) + (0.8396,-0.1)$);
        \coordinate (sw2) at ($(sw1) + (0.8396,0)$);
        \coordinate (sw3) at ($(sw2) + (0.8396,0)$);
        \coordinate (sw4) at ($(sw3) + (0.8396,0)$);
        \coordinate (sw5) at ($(sw4) + (0.8396,0)$);
        \coordinate (sw6) at ($(sw5) + (0.8396,0)$);
        \coordinate (sw7) at ($(sw6) + (0.8396,0)$);
        \coordinate (sw8) at ($(sw7) + (0.8396,0)$);
        \coordinate (sw9) at ($(sw8) + (0.8396,0)$);
        \coordinate (sw10) at ($(sw9) + (0.8396,0)$);
        \coordinate (sw11) at ($(sw10) + (0.8396,0)$);
        \coordinate (sw12) at ($(sw11) + (0.8396,0)$);
        \coordinate (sw13) at ($(sw12) + (0.8396,0)$);
        \coordinate (sw14) at ($(sw13) + (0.8396,0)$);
        \coordinate (sw15) at ($(sw14) + (0.8396,0)$);
        \coordinate (END) at (14.5,-0.2);
        
        % Timeline
        \draw [thick,-stealth] (START) -- (END);
        
        % Ticks at the regular intervals
        \draw [thick] (START) -- ++ (0,-0.1); 
        \draw [thick] (START) -- ++ (0,0.1);
        %\draw [thick] (700) -- ++ (0,0.2);
        \draw [thick] (sw1) -- ++ (0,0.2);
        \draw [thick] (sw2) -- ++ (0,0.2);
        \draw [thick] (sw3) -- ++ (0,0.2);
        \draw [thick] (sw4) -- ++ (0,0.2);
        \draw [thick] (sw5) -- ++ (0,0.2);
        \draw [thick] (sw6) -- ++ (0,0.2);
        \draw [thick] (sw7) -- ++ (0,0.2);
        \draw [thick] (sw8) -- ++ (0,0.2);
        \draw [thick] (sw9) -- ++ (0,0.2);
        \draw [thick] (sw10) -- ++ (0,0.2);
        \draw [thick] (sw11) -- ++ (0,0.2);
        \draw [thick] (sw12) -- ++ (0,0.2);
        \draw [thick] (sw13) -- ++ (0,0.2);
        \draw [thick] (sw14) -- ++ (0,0.2);
        \draw [thick] (sw15) -- ++ (0,0.2);

        % Labels
        \draw (START) ++ (0,-0.1) node[below]{$G1$};
        \draw (sw1) node[below]{$T14$};
        \draw (sw2) node[below]{$L46$};
        \draw (sw3) node[below]{$D6$};
        \draw (sw4) node[below]{$D6$};
        \draw (sw5) node[below]{$D6$};
        \draw (sw6) node[below]{$L45$};
        \draw (sw7) node[below]{$L57$};
        \draw (sw8) node[below]{$T72$};
        \draw (sw9) node[below]{$G2$};
        \draw (sw10) node[below]{$D5$};
        \draw (sw11) node[below]{$D5$};
        \draw (sw12) node[below]{$D5$};
        \draw (sw13) node[below]{$L78$};
        \draw (sw14) node[below]{$D8$};
        \draw (sw15) node[below]{$D8$};
    \end{tikzpicture}
    }
    \caption{Results of DynOPF-R.}
    \label{fig:dyn-res}
\end{figure*}

The output switching order is shown in Table~\ref{tab:9bus-dynamic}, and the transients are depicted in Fig.~\ref{fig:dyn-res}.  The algorithm first picks up all loads at Bus 6, then energizes towards Generator 2.  After picking up Generator 2, 5 loads are picked up. This simulation requires 12.3 hours for the branch-and-bound solver.

% \node[anchor=south west,inner sep=0] at (0,0) {\includegraphics[width=\textwidth]{some_image.jpg}};
% https://tex.stackexchange.com/questions/9559/drawing-on-an-image-with-tikz

\begin{table*}[ht]
  \centering
  \caption{Results of DynOPF-R on IEEE 9-Bus Model.}
    \resizebox{\columnwidth}{!}{
  \begin{tabular}{c|c|c|c|c|c|c|c|c|c|c|c|c|c|c|c|c}
    \toprule
    Time [s] & Init. & 45 & 90 & 135 & 180 & 225 & 270 & 315 & 360 & 405 & 450 & 495 & 540 & 585 & 630 & 675  \\ \midrule 
    Component & G & T & L & D & D & D & L & L & T & G & D & D & D & L & D & D   \\ \midrule
    Buses & 1 & 1--4 & 4--6 & 6 & 6 & 6 & 4--5 & 5--7 & 7--2 & 2 & 5 & 5 & 5 & 7--8 & 8 & 8 \\ \bottomrule
  \end{tabular}
  }
  \label{tab:9bus-dynamic}
\end{table*}

\begin{table*}[ht]
  \centering
  \caption{Results of DynOPF-R on IEEE 9-Bus Model, 2-Island Case.}
  \begin{tabular}{c|c|c|c|c|c|c|c|c|c|c|c|c}
    \toprule
    Time [s] & Init. & 45 & 90 & 135 & 180 & 225 & 270 & 315 & 360 & 405 & 450 & 495  \\ \midrule 
    Component & G & T & T & L & L & D & D & D & L & D & D & D    \\ \midrule
    Buses & 1, 2 & 2--7 & 1--4 & 5--7 & 4--5 & 5 & 5 & 5 & 7--8 & 8 & 8 & 8 \\ \bottomrule
  \end{tabular}
  \label{tab:9bus-dynamic-2island}
\end{table*}

It is observed in Fig.~\ref{fig:dyn-res} that the active power and frequency transients behave as expected.  While the electrical power demanded from the generators increases sharply with a load energization, the generator (turbine and SM) reacts with a TC and offset by a gain.  Frequency deviations from nominal are proportional to the difference between mechanical and electrical power of the generator, by design.  Further, the algorithm sets $\Delta \omega_{\textrm{ref},1} = 0.0207$ and $\Delta \omega_{\textrm{ref},2} = 0.0214$ p.u., causing the starting frequency of the island to be above 50 Hz to avoid trips via underfrequency protection.  Non-nominal frequency setpoints are also recommended in some real restoration plans~\cite{ELIAresto}.

\begin{figure}[t]
    \centering
    \includegraphics[width=8.8cm]{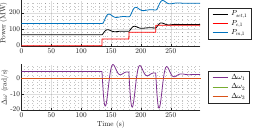}
    \caption{Results of DVLP with load energizations at Bus 5.}
    \label{fig:dvlp-res}
\end{figure}

It is notable that, in contrast with the static formulation, the algorithm picks up loads at Bus 6 before loads at Bus 5.  To show why this occurs, a Dynamic Validation LP (DVLP) is constructed with the power balance and dynamic constraints of the full formulation, but excluding the switching constraints.  Then, a series of switches is given to the DVLP as parameters.  For this case study, we apply three load energizations at Bus 5 after the energization of the transformer 1--4 and line 4--5 with a time horizon of 300 seconds, and the lower frequency limit is relaxed to $\Delta f_{min} = -3$ Hz.  The results are shown in Fig.~\ref{fig:dvlp-res}.  It can be seen that the frequency transient for this switching sequence does not satisfy the previously set threshold of $-2$ Hz, and confirmed by the fact that the DVLP with  $\Delta f_{min} = -2$ is infeasible.  This analysis illustrates the necessity of the dynamic formulation: without considering frequency constraints, the optimized switching sequence can be practically infeasible.

The DVLP can be further used to validate the optimality of the dynamic setpoints, given the optimized energization order as parameters.  Indeed, the optimal frequency setpoints computed using DVLP with frequency range [48, 51.5] Hz are $\Delta \omega_\textrm{ref} =$ 0.0207 and 0.0214 p.u. for G1 and G2, respectively.

Finally, for the proposed formulation to be practically applicable considering existing island-based restoration strategies~\cite{spain-bo-report}, it is prudent to further verify its functionality using more than one initially energized island.  To this end, we execute a second Case Study using the same parameters as listed in Table~\ref{tab:opt-param}, but with $T_{max} = 9$ minutes.  Further, we initialize the system with Generators 1 and 2 energized.  The purpose of this test is to ensure that the formulation of the power angle constraints is robust to island synchronization manoeuvres.  The resulting energization order for this case is shown in Table~\ref{tab:9bus-dynamic-2island}.  This case requires 7.2 hours of convergence time.  The synchronization between the two islands occurs with the energization of the line between buses 4 and 5 at $t=180$ seconds, allowing the loads to be energized with a stronger island containing two generators.

\section{Conclusions}\label{chap:conc}
In restoration plans of TSOs, it is common either to use heuristics to account for dynamic constraints in the restoration process, or to conduct detailed studies only on a defined restoration path. In this paper, we propose a formulation of the optimal PSR problem which considers frequency dynamics, inspired by the synchronous machine equations.  The approach relies on DC OPF, well-suited for application to high-voltage transmission networks, and by extension, restoration planning processes of TSOs.  We validate the static formulation of the switching constraints for global optimality using a tree-search method. A dynamic formulation of the restoration problem is proposed which includes both static and transient droop components of synchronous machine governor dynamics.  It is shown via the application of the formulation to the IEEE 9-Bus model that the switching sequence optimized using the static PSR problem formulation (OPF-R) would violate the chosen frequency tolerance band of [48, 51.5] Hz for DynOPF-R, illustrating the importance of considering frequency dynamics in restoration planning processes.

In future work, solutions to improve the scalability of the method can be explored, as well as its extension to include voltage and reactive power constraints.  The impacts of modern energy storage systems on the restoration process can additionally be further explored.

\bibliographystyle{IEEEtran}
\bibliography{bibliography}

\end{document}